\documentclass[fleqn,usenatbib]{mnras}
\usepackage{lineno}
\usepackage{tabularx}
\usepackage{graphicx}
\usepackage[T1]{fontenc}
\usepackage{ae,aecompl}
\usepackage{booktabs}
\usepackage{gensymb}
\usepackage{hyperref}

\usepackage{newtxtext,newtxmath}
\usepackage{amsmath}
\usepackage{natbib}
\usepackage{afterpage}
\usepackage{tablefootnote}
\usepackage{threeparttablex}
\usepackage{bm}
\usepackage{multicol}
\usepackage{pdflscape}
\usepackage{arydshln}
\usepackage{kantlipsum}

\def\ltsim{\mathrel{\hbox{\rlap{\hbox{\lower3pt\hbox{$\sim$}}}\hbox{\raise2pt\hbox{$<$}}}}}
\def\gtrsim{\mathrel{\hbox{\rlap{\hbox{\lower3pt\hbox{$\sim$}}}\hbox{\raise2pt\hbox{$>$}}}}}

\title[X-ray Spectral Variability of TeV Blazars]{Sub-Day Timescale X-ray Spectral Variability of the TeV Blazars Mrk 421 and 1ES 1959+650}
\author[Das \& Chatterjee]{
Susmita Das,$^{1}$\thanks{E-mail: susmita.rs@presiuniv.ac.in}
Ritaban Chatterjee$^{1}$
\\
$^{1}$School of Astrophysics, Presidency University, 86/1 College Street, Kolkata, West Bengal, India 700073.
}

\begin{document}
\label{firstpage}
\pagerange{\pageref{firstpage}--\pageref{lastpage}}
\maketitle
\begin{abstract}
We present X-ray spectra ($0.7-20$ keV) of two high synchrotron-peaked blazars Mrk 421 and 1ES 1959+650 from simultaneous observations by the SXT and LAXPC instruments onboard \textit{AstroSat} and the \textit{Swift}-XRT during multiple intervals in 2016-19. The spectra of individual epochs are satisfactorily fitted by the log-parabola model. We carry out time-resolved X-ray spectroscopy using the \textit{AstroSat} data with a time resolution of $\sim$10 ks at all epochs, and study the temporal evolution of the best-fit spectral parameters of the log-parabola model. The energy light curves, with duration ranging from $0.5-5$ days, show intra-day variability and change in brightness states from one epoch to another. We find that the variation of the spectral index ($\alpha$) at hours to days timescale has an inverse relation with the energy flux and the peak energy of the spectrum, which indicates a harder-when-brighter trend in the blazars. The variation of curvature ($\beta$) does not follows a clear trend with the flux and has an anti-correlation with $\alpha$. Comparison with spectral variation simulated using a theoretical model of time variable nonthermal emission from blazar jets shows that radiative cooling and gradual acceleration of emitting particles belonging to an initial simple power-law energy distribution can reproduce most of the variability patterns of the spectral parameters at sub-day timescales.
\end{abstract}


\section{Introduction}
An active galactic nucleus (AGN) radiates an enormous amount of energy, which may sometimes exceed the host galaxy luminosity. This huge energy is supplied by the accretion of matter towards the central super-massive black hole (SMBH) and is radiated through different components of the AGN, e.g., accretion disc, emission line clouds, dusty torus and the outflow of bipolar jets containing magnetized plasma moving at speed close to that of light. When the axis of the relativistic jet is very close to the observer's line of sight, the AGN is classified as a blazar \citep{bla78,urr95}. It is one of the subclasses of radio-loud AGN, in which the jet emission, amplified by relativistic beaming, dominates the entire spectral energy distribution, which often extends from radio to $\gamma$-rays \citep[e.g.,][]{ghise1993, giommi12}. Consequently, we can see dramatic variation of luminosity at minutes-years timescale \citep[e.g.,][]{abdo10a,bonning12,chatt12,massaro15,marscher2016,rajput20}. The broadband continuum emission of blazars is defined by two humps at lower and higher energy regimes. At low energy, the peak is at IR-optical sometimes extending to UV/X-ray, and the higher-energy peak lies at the GeV band.

The origin of the two broad peaks in the spectral energy distribution (SED) can be explained by the radiation mechanisms that are responsible for jet emission. In the presence of magnetic field, the relativistic electrons in the jet generate synchrotron emission that dominates the lower energy peak of the SED \citep{bregman1981,urry1982,impey1988,marscher1998,ghis17}. On the other hand, the origin of the higher-energy emission has two possible scenarios. The jet plasma may be made of electrons, positrons, and protons. In the so-called ``leptonic'' model, protons are absent or their contribution to jet emission is negligible. In that scenario, inverse-Compton (IC) scattering by the electrons contributes to the higher energy emission by the up-scattering of lower energy ``seed'' photons, which may be the synchrotron photons in the jet itself \citep{koni1981,maras1992,sikora1994,bloom1996,mast1997,bott07} or photons external to the jet, e.g., from the dusty torus or the broad line region \citep{dermer1992,ghise1998,blejo2000,bott10} which are defined as synchrotron self-Compton (SSC) or external Compton (EC) process, respectively. On the other hand, protons contribute significantly to the jet emission in the ``hadronic'' model. In that case, relativistic protons in the jet may emit synchrotron radiation, or they may produce secondary particles via their interaction with photons \citep{mucke01, mucke03, bottcher13,botta16,acker16}, which contribute to the $\gamma$-ray emission of blazars. 

Significant fluctuation of the total flux over a range of timescales is a characteristic property of blazars and their spectral nature, e.g., the frequency of the peak and shape of the lower and higher energy humps vary along with the total flux. In this work, we focus on the X-ray spectral variation of two blazars, namely, Mrk 421 and 1ES 1959+650, at minutes to days timescales. Both of those blazars are high synchrotron peaked (HSP), i.e., the lower energy peak of their SED is at the UV-X-ray energy range \citep[$\nu_{sync}>10^{15}$ Hz;][]{beckmann02,abdo11,ban19,magic20}. Therefore, X-ray emission provides information about the highest energy electrons in the jet. 

Mrk 421 is the first ever TeV-detected blazar in a nearby galaxy at $z=0.031$. It was first observed by EGERT and Whipple Observatory in 1992 \citep{lin1992,punch1992,piner1999}. Later, the broadband SED and the underlying physical processes in Mrk 421 have been studied using numerous multi-wavelength observations \citep[e.g.,][]{brinkmann05,lichti08,fossati08,horan09,abdo11,marko22,albert22}. The lower energy (synchrotron) peak in the SED of Mrk 421 is at $0.1-10$ keV and the higher energy peak is at $\sim 50$ GeV \citep{abdo11,ban19}. The spectrum of Mrk 421 at the X-ray energies shows a significant curvature in its shape \citep{landau1986,sambruna1996,inoue1996,takahashi1996}. The curved spectrum, at different flux states, has been satisfactorily fitted with a log-parabola model \citep{tanihata04,massaro04,tramacere07,massaro08,tramacere09,chen14,kapa17,bhatta18,kapa20} as well as a broken power-law model \citep{takahashi1996,brinkmann01,brinkmann05,chen14}. In addition, using BeppoSAX, Swift, and XMM-Newton observations, \citet{massaro04,massaro06} showed that the electron energy distribution responsible for the X-ray emission through synchrotron radiation in Mrk 421 approximately follows a log-parabolic function. \citet{massaro08} showed that the X-ray spectra of several other HSP blazars are well described by the log-parabola model and exhibit significant correlation between certain spectral parameters, which is consistent with the statistical/stochastic acceleration process in the jet. The curvature of the X-ray spectrum is also well described by certain physically motivated models of energy-dependent particle distributions, e.g., energy-dependent diffusion (EDD), energy-dependent acceleration (EDA), power law with a high-energy cut-off \citep{goswami18,goswami20,hota21,rukaiya22}. 

1ES 1959+650 ($z=0.048$) is one of the first blazars detected at TeV energies by the Utah Seven Telescope Array experiment \citep{nishiyama1999}. It was studied at X-ray energies for the first time by the Einstein-IPC in 1992 \citep{elvis1992}, and later by ROSAT in 1996, BeppoSAX in 1997 and by Rossi X-Ray Timing Explorer (RXTE) in 2000 \citep{beckmann02,giebels02}. The broadband SED has been studied by many authors \citep[e.g.,][]{bott10,aliu13,kapa16,magic20,chandra21} and it is found that both leptonic and hadronic models may be used to fit the SED satisfactorily. At the X-ray band, the spectrum is curved and is fit better by the log-parabola function compared to other models \citep{tagli03,tramacere09,kapa18,chandra21,wani23}. From X-ray spectral fitting, it is possible to probe relations, if any, between the best-fit model parameters, which may provide physical information regarding the emission processes involved. In \citet{zahir21}, the X-ray spectrum is described by synchrotron radiation generated by different particle energy distributions, e.g., log-parabola and broken power-law, with the latter providing a better fit. They have shown that the particle number density is anti-correlated with the best-fit power-law index below the break energy, which depicts a connection of acceleration and cooling mechanisms of the particles in the emission region. 

Polarization at multiple wave bands provides information, which are complimentary to that obtained from variability and spectroscopy. Hence, it is an effective probe of the physical processes in jets. In the last two years, Imaging X-ray Polarimetry Explorer (IXPE) is carrying out X-ray polarization observations of blazars with unprecedented details \citep{weiss2022}. Strong X-ray polarization ($\sim$10-20\%) has been detected in several HSP blazars, including those in our sample, e.g., Mrk 421 and 1ES 1959+650, while in low synchrotron peaked (LSP) blazars IXPE observations have mostly come up with only upper limits but no detection \citep[e.g.,][]{liodakis22Natur,laura23Nat,middei23ApJ}. That has provided support to the notion that X-rays generated in the synchrotron process are strongly polarized while it is weak in X-rays produced by the inverse-Compton process, latter being the case in LSP blazars. In the hadronic scenario, X-rays in LSP blazars may be due to proton synchrotron process and expected to be strongly polarized. Therefore, non-detection of polarization in those sources indicates that the hadronic process does not contribute significantly to the X-ray emission in those blazars \citep{peirson2018,peirson2019,manel24ApJ,kim24A&A}. The detected X-ray polarization fraction in the above HSP blazars is $\sim$1.5-3 times higher than that at the optical and longer wavelengths. That is consistent with the energy stratification of blazar jets, i.e., highest energy particles, such as those emitting X-rays in HSP blazars, remain confined to a small region, e.g., behind a moving shock front that is causing the energization of the emitting particles while lower energy particles, emitting at longer wavelengths, are located in a larger region. Magnetic field in the smaller region containing X-ray emitting particles is stronger and more aligned giving rise to higher polarization fraction \citep{liodakis22Natur,kouch24A&A,alan24Galax}. 

The two blazars in our sample have been observed with \textit{AstroSat} at several epochs during 2016-19 giving rise to X-ray light curves and spectra at minutes to days as well as $\sim$months timescales. The temporal variability of the soft and hard X-ray flux of the two blazars, during those epochs, was analysed and interpreted in a previous work \citep{das23}. They found that the hard and soft X-ray variability are strongly correlated and the time lag is zero in the majority of the epochs although significant hard and soft lags were also present in a few cases. Here, we aim to find the best-fit model of the X-ray spectra during those epochs and whether they tend to fluctuate from one epoch to another. We further study the time-resolved spectra to study the variability at sub-day timescales and find correlation between the best-fit spectral parameters. We explore the possible physical explanation of the observed correlation trends. For that purpose, we compare the results with those of a theoretical model consisting of a multi-zone emission region, in which initially quiescent electrons are energized by a moving shock front. The electron energy distribution and magnetic field in different cells in the emission region are independent of each other and the former evolves with time due to emission of nonthermal radiation. Such an emission scenario in blazar jets is consistent with various observations, including those by IXPE as discussed above. The time resolved spectra obtained from the above model and the correlation of the best-fit model parameters may be compared with the observed results obtained from analyzing the \textit{AstroSat} data in order to test the said emission scenario and to constrain the geometric and emission parameters in the jets of those blazars.

In Section 2, we describe the data reduction process to generate the X-ray spectra from \textit{AstroSat} and \textit{Swift} observations. In Section 3, we analyse the $0.7-20$ keV spectrum for each epoch and perform time-resolved spectroscopy in bins of a few hours to study spectral variation with time. In Section 4, we discuss our results and present a theoretical model of non-thermal emission from blazar jets, and use it to interpret the spectral variability at hours to days timescales. Finally, in Section 5, we summarise our findings. 

\begin{figure}
\centering
\includegraphics[height=11cm,width=\columnwidth]{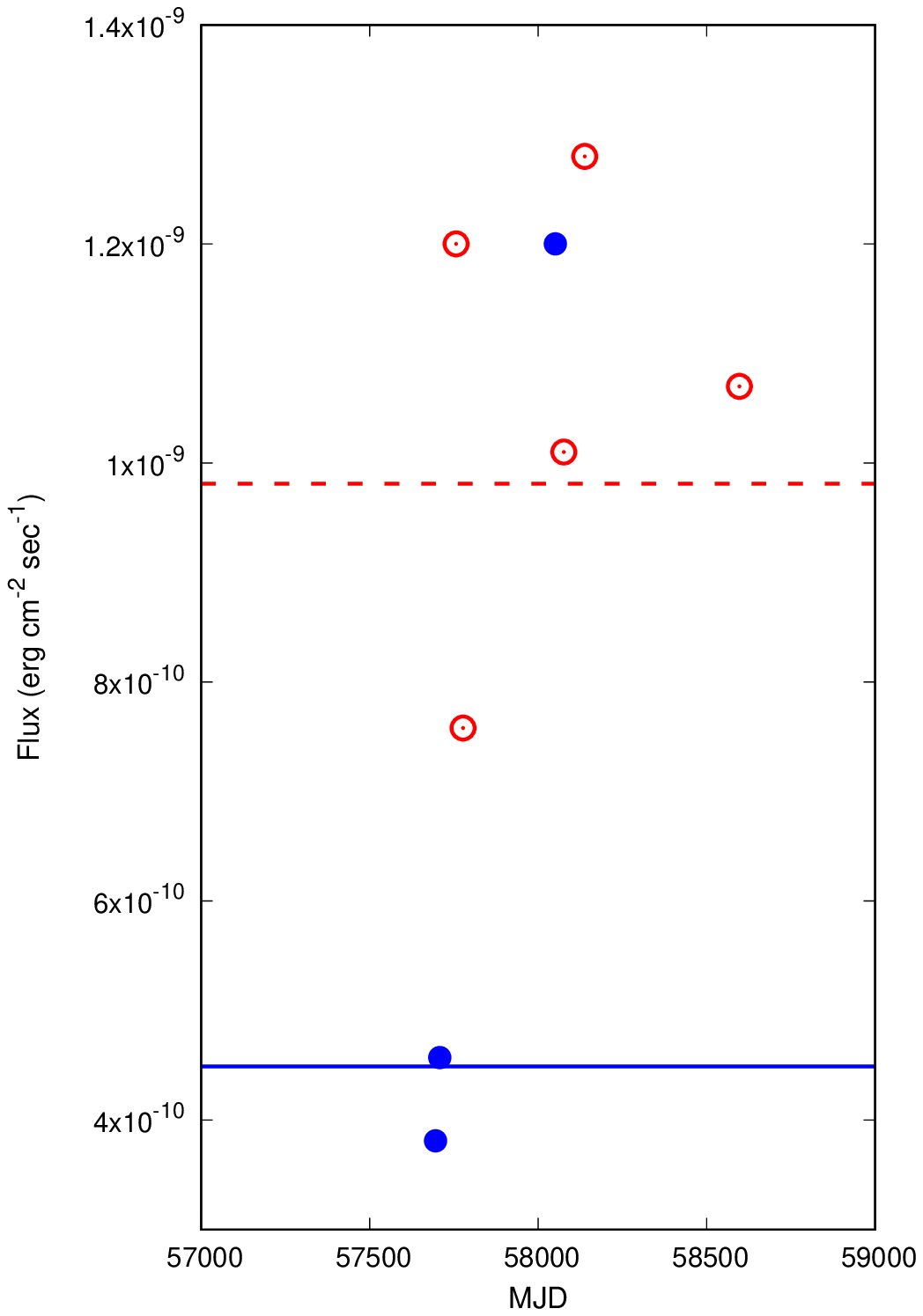} 
\caption{Brightness state of Mrk 421 (red) and 1ES 1959+650 (blue) where circles denote the epochs of \textit{AstroSat} observations and the lines are average monitored flux from \textit{Swift}-XRT during 2013-23.}\label{flux_state}
\end{figure}

\begin{figure*}
\centering
\includegraphics[height=20cm,width=\textwidth]{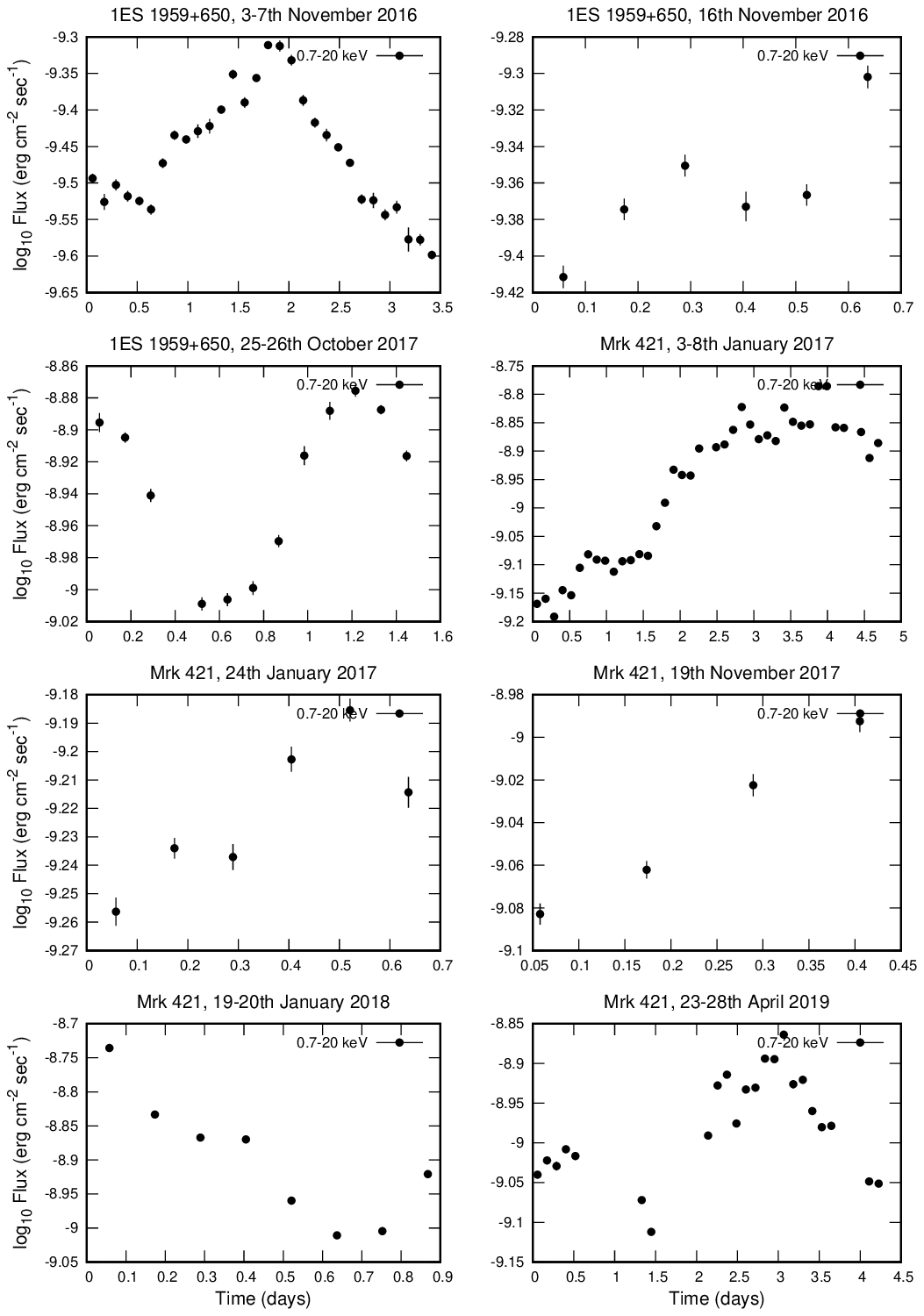} 
\caption{The energy flux light curves of the blazars 1ES 1959+650 and Mrk 421 from time-resolved spectroscopy with 10 ks bins obtained from fitting the \textit{AstroSat} data with log-parabola model at multiple epochs during 2016-19.}\label{flux_lc}
\end{figure*}

\section{Data}
The X-ray data of the blazars Mrk 421 and 1ES 1959+650 used in this work are from \textit{AstroSat} and the \textit{Neil Gehrels Swift Observatory}. We use data from the instruments named Soft X-ray Telescope \citep[SXT;][]{kpsing14,kpsing16,kpsing17} and Large Area X-ray Proportional Counter \citep[LAXPC;][]{yadav16,antia17} of \textit{AstroSat} and X-Ray Telescope (XRT) of \textit{Swift}, working in the energy range $0.3-8$, $3-80$, and $0.3-10$ keV, respectively. These observed data sets are available in the respective  repositories\footnote{\url{https://astrobrowse.issdc.gov.in/astro_archive/archive/Home.jsp}}$^,$\footnote{\url{https://heasarc.gsfc.nasa.gov/}}. The data consist of three epochs for 1ES 1959+650 and five for Mrk 421 during 2016-19 as given in Table \ref{logpar_para}. Preliminary reduction and analyses of SXT and LAXPC data, and details of the final data sets are described in \citet{das23}.

We have collected \textit{Swift}-XRT data, the duration of which overlaps with \textit{AstroSat} observations during 2016-19. We get a total of eight observations of Mrk 421 and four observations of 1ES 1959+650 of duration $\sim$1 ks in Windowed Timing (WT) mode.
The light curves and spectra of the source at $0.3-10$ keV energy range are generated by the \textit{Swift}-XRT data products building tool\footnote{\url{https://www.swift.ac.uk/user_objects/index.php}}, which uses \textit{HEASoft} version 6.29 \citep{evans09}. Figure \ref{flux_state} exhibits the brightness levels of the two blazars during the epochs used here compared to their respective average flux during 2013-2023 using \textit{Swift}-XRT data.

We note that X-ray-bright HSP blazars, such as the two in our sample, have been observed multiple times by several X-ray telescopes, and their spectra have been analysed and interpreted in detail. However, \textit{AstroSat} has observed the blazars Mrk 421 and 1ES 1959+650 several times within a short span of a few years. Individual observations are able to provide data points with sufficient S/N ratio even when binned to 10-minute intervals in some cases \citep{das23} while multiple epochs of observations provide variability information at months-years timescales. Such detailed flux and spectral variability information at a broad (0.7-20 keV) energy range is not easily available even for bright HSP blazars like the above two. Therefore, the dataset used here is uniquely comprehensive.

\begin{table*}
\centering
\caption{Best-fit parameter values from log-parabola modeling of XRT + SXT + LAXPC spectra.}\label{logpar_para}
\begin{tabular}{|l|l|l|l|l|l|l|l|} 
\hline
Source & Epoch & Constant$^{*}$ &$\alpha$ & $\beta$ & $E_{peak}$ (keV) & log$_{10}$Flux in CGS &$\chi^2/dof$  \\
 & & & & & & ~~ ($0.7-20$ keV) & \\
\hline
1ES 1959+650 & 2016, 3-7th Nov & 0.77 $\pm$ 0.02 & 1.99 $\pm$ 0.01 & 0.35 $\pm$ 0.02 & 0.98 $\pm$ 0.05 & -9.42 $\pm$ 0.15 & 1.07 \\
 & 2016, 16th Nov & 0.81 $\pm$ 0.03 &1.97 $\pm$ 0.03 & 0.42 $\pm$ 0.03 & 1.09 $\pm$ 0.09 & -9.34 $\pm$ 0.006 & 1.03 \\
 &  2017, 25-26th Oct & 0.77 $\pm$ 0.01 & 1.72 $\pm$ 0.01 & 0.32 $\pm$ 0.01 & 2.69 $\pm$ 0.07 & -8.92 $\pm$ 0.12 & 1.17 \\ 
 \hline
 Mrk 421 & 2017, 3-8th Jan & 0.71 $\pm$ 0.01 & 1.81 $\pm$ 0.01 & 0.26 $\pm$ 0.02 & 2.25 $\pm$ 0.05 & -8.92 $\pm$ 0.08 & 1.4\\
 & 2017, 24th Jan & 0.71 $\pm$ 0.02 & 1.97 $\pm$ 0.02 & 0.33 $\pm$ 0.02 & 1.07 $\pm$ 0.07 & -9.12 $\pm$ 0.01 & 1.11 \\
 & 2017, 19th Nov & 0.77 $\pm$ 0.02 & 1.91 $\pm$ 0.02 & 0.37 $\pm$ 0.02 & 1.30 $\pm$ 0.08 & -9.00 $\pm$ 0.003 & 0.95 \\
 & 2018, 19-20th Jan & 0.75 $\pm$ 0.01 & 2.02 $\pm$ 0.01 & 0.31 $\pm$ 0.01 & 0.91 $\pm$ 0.05 & -8.89 $\pm$ 0.02 & 1.38 \\
 & 2019, 23-28th April & 0.69 $\pm$ 0.01 & 1.94 $\pm$ 0.01 & 0.12 $\pm$ 0.01 & 1.78 $\pm$ 0.12 & -8.97 $\pm$ 0.04 & 1.44 \\
\hline
\end{tabular}
\begin{threeparttable}
 \begin{tablenotes}
        \item [*]Constant is the flux normalization factor of LAXPC corresponding to SXT.
 \end{tablenotes}
 \end{threeparttable}
\end{table*}

\section{Results}
\subsection{X-ray Spectral Analyses}
 We carry out the spectral analysis using \texttt{XSPEC} (version 12.10), a tool of \textit{HEASoft}. The command \texttt{GRPPHA} helps to group the source spectrum with response and background files. The energy channels are grouped here by a minimum of 30 counts per bin. 
 We analyse SXT and LAXPC spectra jointly in the energy range $0.7-20$ keV by loading the spectra in \texttt{XSPEC}. Below 0.7 keV, there are certain instrumental artifacts in the SXT spectra while the background, in LAXPC, starts to dominate at energies higher than $20$ keV \citep{zahir21,hota21}. Thus we ignore those energy ranges to carry out our analyses. The spectral data from \textit{Swift}-XRT during the epochs 2017 Jan 3 and 2019 Apr 23 for Mrk 421, and 2016 Nov 3 and 2017 Oct 25 for 1ES 1959+650 at $0.7-7$ keV range are used because those epochs overlap with that of the \textit{AstroSat} observations. 

 The combined spectra are fitted using an empirical model, namely, log-parabola in \texttt{XSPEC}. There is a flux normalization offset between different instruments, which is represented by a constant factor in the spectral modeling. The Galactic hydrogen column density ($n_H$) values for the line of sight containing Mrk 421 and 1ES 1959+650 are fixed at $1.92 \times 10^{20}$ cm$^{-2}$ and $1.07 \times 10^{21}$ cm$^{-2}$ \citep{kalberla05}, respectively. We add $3\%$ systematic error and perform the gain correction for SXT spectra \citep{chandra21}.
The time-averaged spectra are satisfactorily fitted by the log-parabola model. The best-fit models and data are shown in the Appendix. 
The best-fit values of the parameters of the model and the reduced $\chi^2$ values are given in Table \ref{logpar_para}. The average flux varies by a factor of $3-4$ between epochs, and the best-fit values of the spectral parameters fluctuate as well. The peak energy shifts from $0.9$ to $2.7$ keV for the two sources. The fit statistics for the epochs described here are comparable to those in the case of the broken power-law model \citet{das23}. We used the former model for the following time-resolved spectral analyses because the synchrotron peak of those two blazars are in the $0.7-20$ keV energy range and log-parabola function is a more natural choice as the shape of the peak. In addition, it is often used by other authors, making it more useful in comparing our results with those found in the literature. 

\subsection{Time-Resolved Spectroscopy}
\begin{figure*}
\centering
\includegraphics[height=18cm,width=\textwidth]{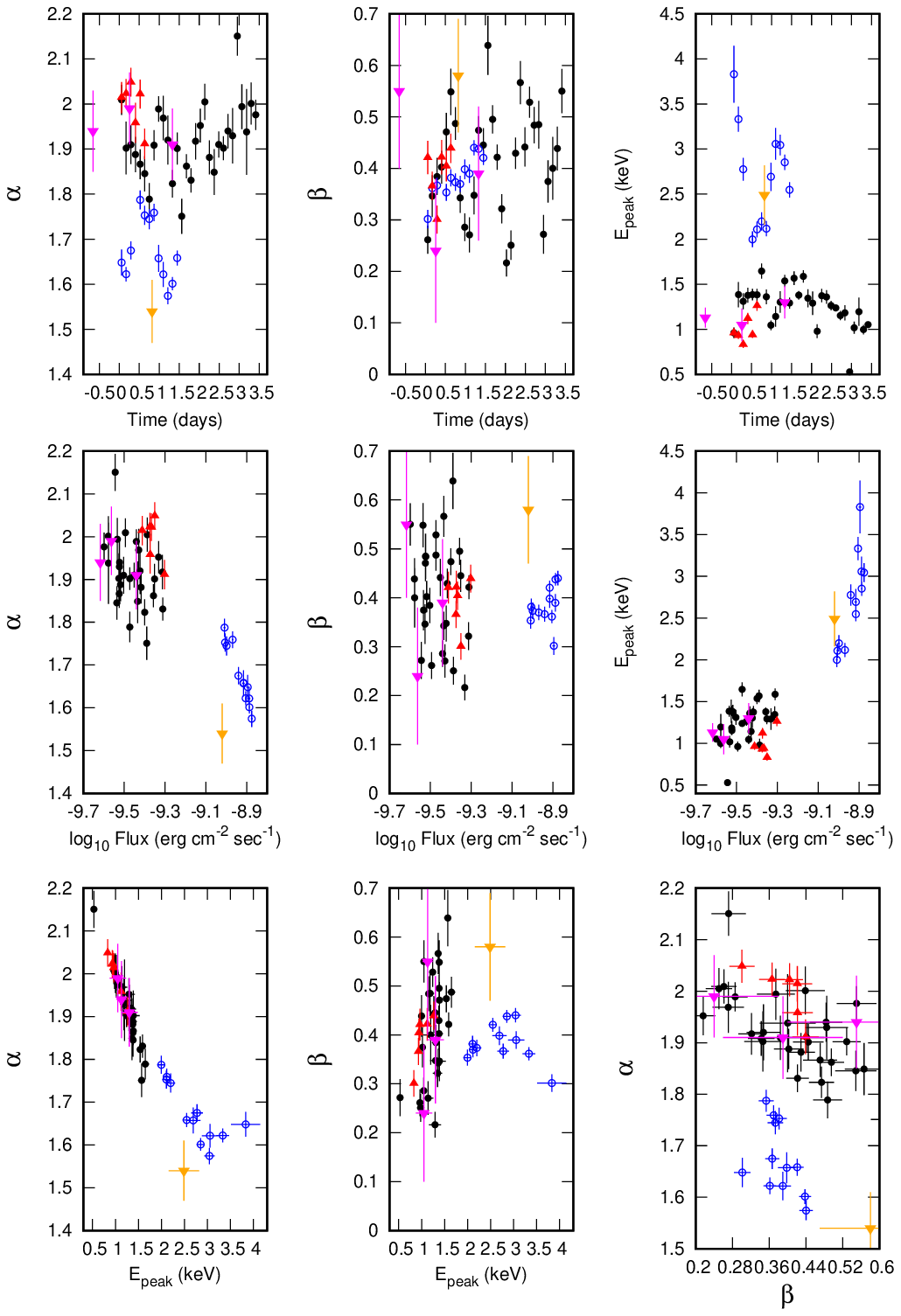} 
\caption{Correlation between best-fit spectral parameters of the log-parabola model from time-resolved spectroscopy of the \textit{AstroSat} data of the blazar 1ES 1959+650 with time bins of length 10 ks. Black solid circles, red triangles and blue open circles denote the 2016 Nov 3, 2016 Nov 16, and 2017 Oct 25 epochs, respectively. The inverted triangles (orange and magenta for 2016 Nov 3 and 2017 Oct 25 epoch respectively) denote \textit{Swift}-XRT observations.}\label{para_log_1959}
\end{figure*}

\begin{figure*}
\centering
\includegraphics[height=18cm,width=\textwidth]{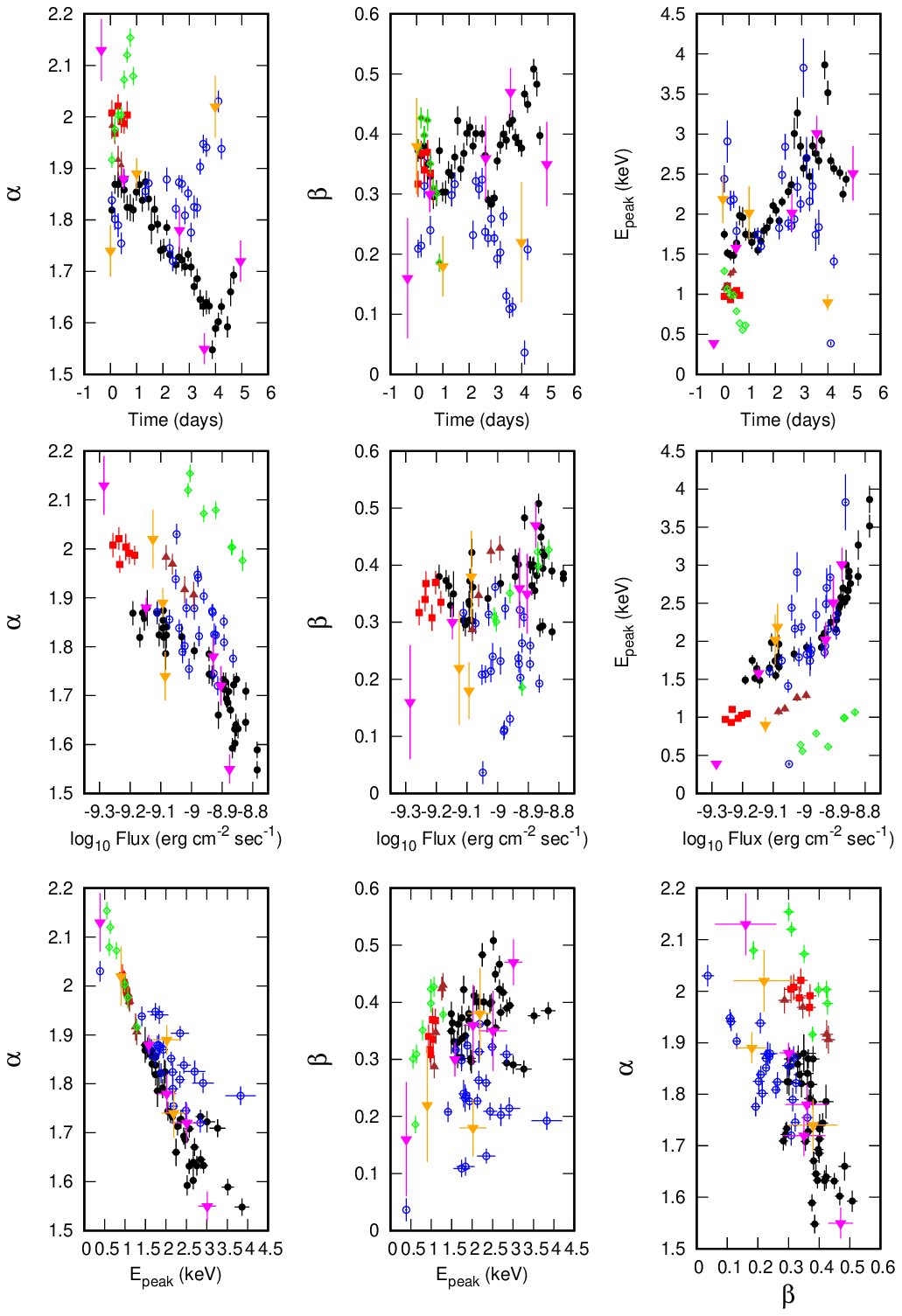} 
\caption{Correlation between best-fit spectral parameters of the log-parabola model from time-resolved spectroscopy of the \textit{AstroSat} data of the blazar Mrk 421 with time bins of length 10 ks. Black solid circles, red filled squares, maroon triangles, green diamonds, and blue open circles denote the 2017 Jan 3, 2017 Jan 24, 2017 Nov 19, 2018 Jan 19, and 2019 Apr 23 epochs, respectively. The inverted triangles (orange and magenta for 2017 Jan 3 and 2019 Apr 23 epoch respectively) denote \textit{Swift}-XRT observations.}\label{para_log_421}
\end{figure*}

To study the nature of spectral variability at sub-day timescale, we divide the data of all epochs of the two blazars into time segments of length $10$ ks. For each time bin, we generate the SXT spectrum using ``filter time file'' in \texttt{XSELECT}\footnote{\url{https://github.com/scsaao/SXT_std_ProductMaker}} and the LAXPC spectrum using the LAXPC software by dividing the good time intervals into the above time bins. The spectra in each bin from SXT and LAXPC are fit together using \texttt{XSPEC} by the log-parabola model within $0.7-20$ keV. We estimate the unabsorbed flux values at $0.7-20$ keV from the fitted time-resolved spectra and determine the energy flux light curves, which are shown in Figure \ref{flux_lc}. The nature of variation exhibited by the energy light curves are similar to that of the photon flux shown in \citet{das23}. We obtain the best-fit values of the spectral model parameters in every time bin of each epoch and then investigate whether there is any inter-relation between the variability of different pairs of parameters. The inter-relations between the derived spectral parameters are shown in Figures \ref{para_log_1959} and \ref{para_log_421}. Similar analyses for the \textit{Swift}-XRT spectra of individual observations overlapping with the \textit{AstroSat} epochs have been carried out and are shown in the same figures.
 
From time-resolved spectroscopy, it is observed that the X-ray spectra of both blazars tend to vary in all of the epochs, i.e., the spectral parameters $\alpha$, $\beta$ and corresponding peak energy ($E_{peak}$) in the log-parabola model are varying significantly from one time-bin to another. 
The spectral index is between $1.5-2.2$ and the curvature range is $0.1-0.7$ which is slightly higher than that reported by \citet{massaro04} but are consistent with the values obtained by \citet{massaro11}. The peak energy varies from approximately 0.5 to 4 keV among all epochs.
The dependence of parameters on flux values shows that $\alpha$ is decreasing and $E_{peak}$ is increasing with increasing flux for all observations, but there is no clear pattern in the variation of $\beta$. On the other hand, $\alpha$ has an inverse relation with $E_{peak}$ while $\beta$ shows a weak correlation with the same, and $\alpha$ and $\beta$ have a trend of anti-correlation with each other. We note that among the observed intervals, the above trends are weaker in both sources during the epochs of lower flux. Table \ref{spearman_rank} shows the strength of correlation and their significance during those of the above epochs, which have durations longer than a day.

\begin{figure*}
   \begin{minipage}{0.45\textwidth}
     \centering
     \includegraphics[width=\columnwidth]{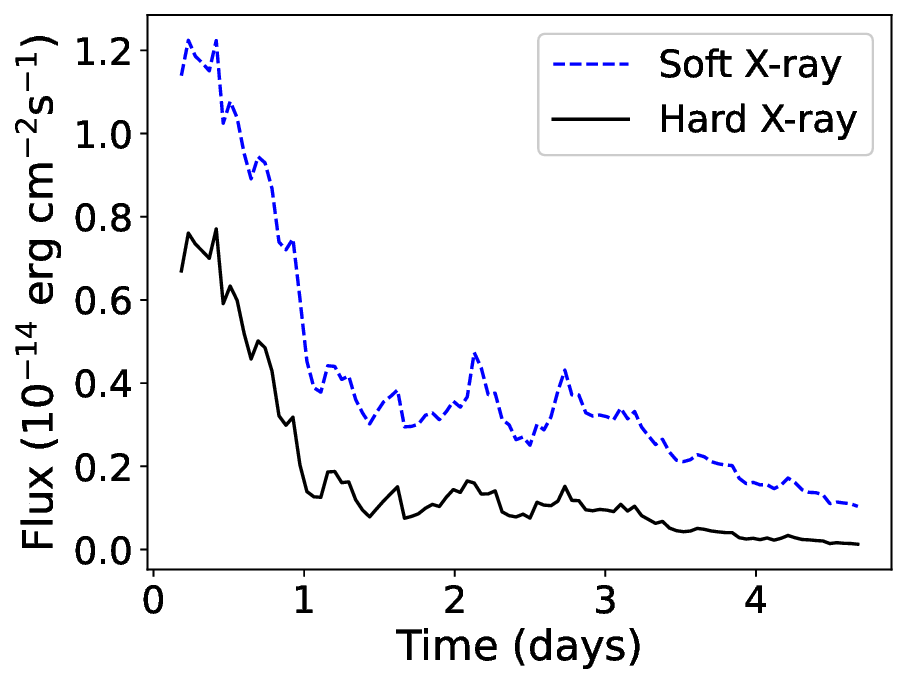}
     \end{minipage}\hfill
   \begin{minipage}{0.45\textwidth}
     \centering
     \includegraphics[width=\columnwidth]{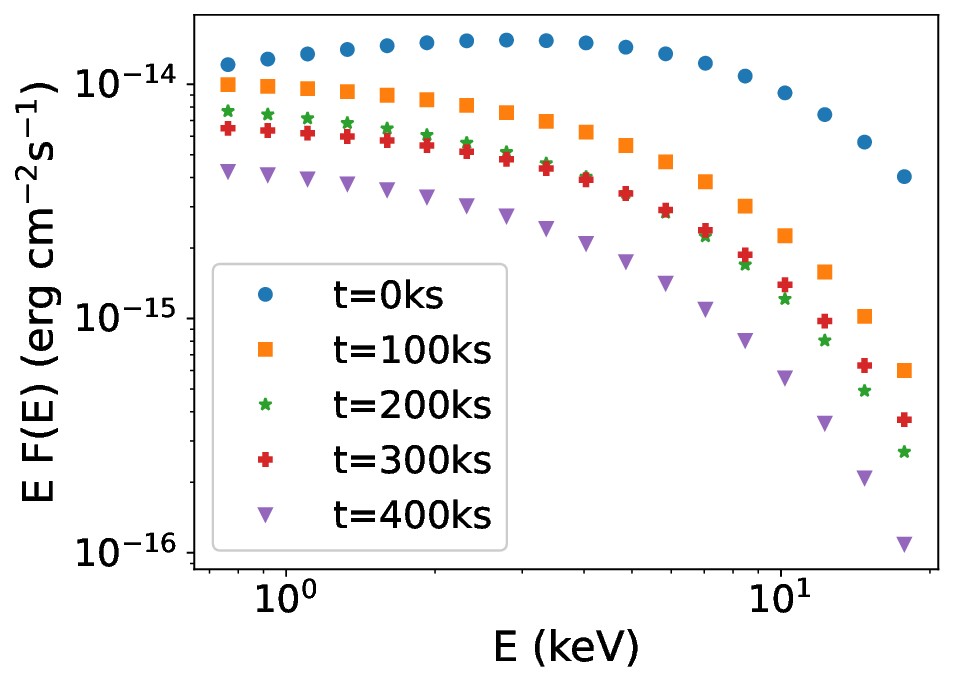}
     \end{minipage}
     \caption{\textit{Left Panel:} Soft ($0.7-7$ keV) and hard X-ray ($7-20$ keV) light curves simulated using the theoretical model discussed in the text for the case of radiative cooling only. \textit{Right Panel:} Evolution of the simulated SED at the X-ray (0.7-20 keV) energy range during a total time interval of $\sim$5 days (comparable to the longer of the observed light curves in our data).}\label{sim_lc_sed_noacc}
\end{figure*}

\begin{figure*}
\centering
\includegraphics[height=17cm,width=\textwidth]{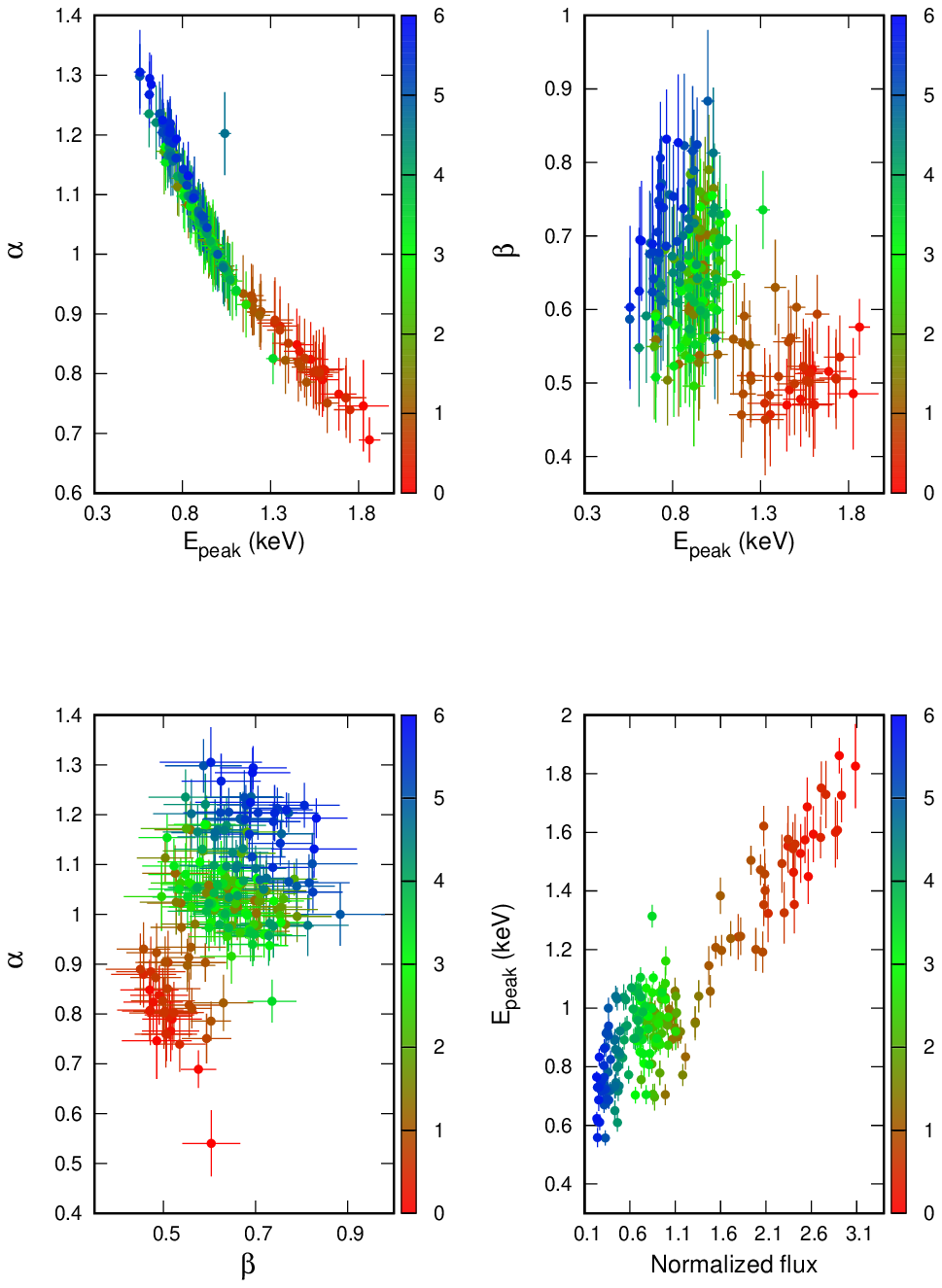} 
\caption{Correlation between the spectral parameters of the log-parabola models, which are best-fit to the simulated X-ray spectra at each snapshot of 5 ks duration for the case of radiative cooling only. The color bar indicates time evolution in days.}\label{sim_parameters_noacc}
\end{figure*}

\begin{figure*}
   \begin{minipage}{0.45\textwidth}
     \centering
     \includegraphics[width=\columnwidth]{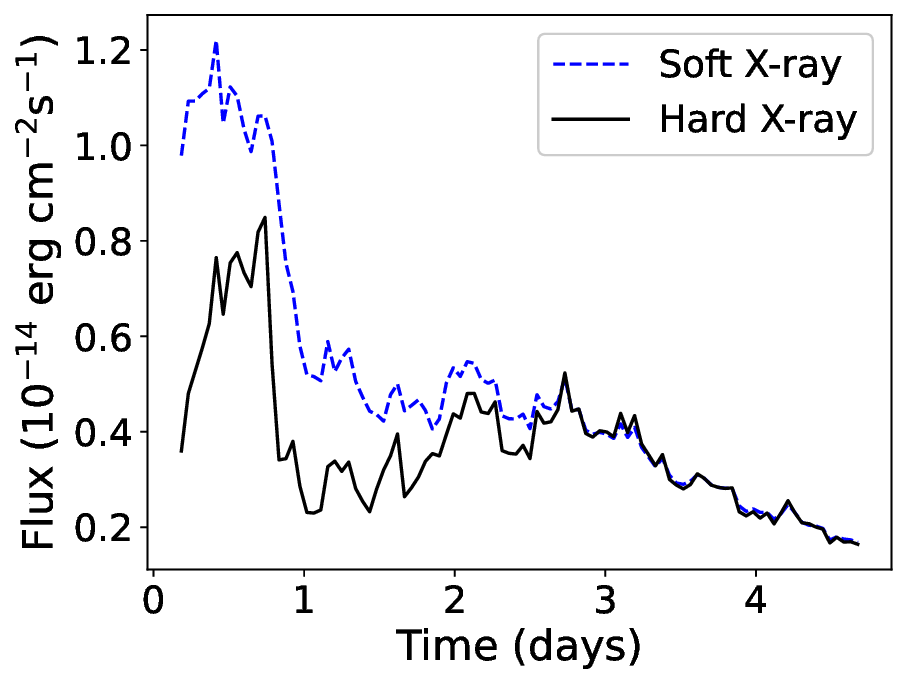}
     \end{minipage}\hfill
   \begin{minipage}{0.45\textwidth}
     \centering
     \includegraphics[width=\columnwidth]{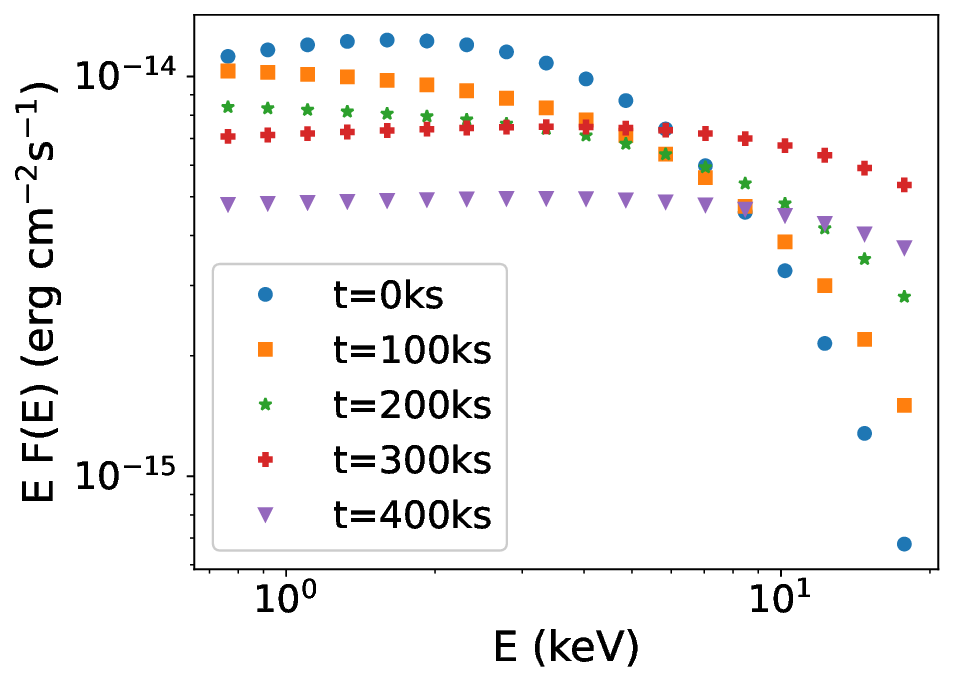}
     \end{minipage}
     \caption{\textit{Left Panel:} Soft ($0.7-7$ keV) and hard X-ray ($7-20$ keV) light curves simulated using the theoretical model discussed in the text for the case of gradual acceleration. \textit{Right Panel:} Evolution of the simulated SED at the X-ray (0.7-20 keV) energy range during a total time interval $\sim 5$ days (comparable to the longer of the observed light curves in our data).}\label{sim_lc_sed_acc}
\end{figure*}

\begin{figure*}
\centering
\includegraphics[height=17cm,width=\textwidth]{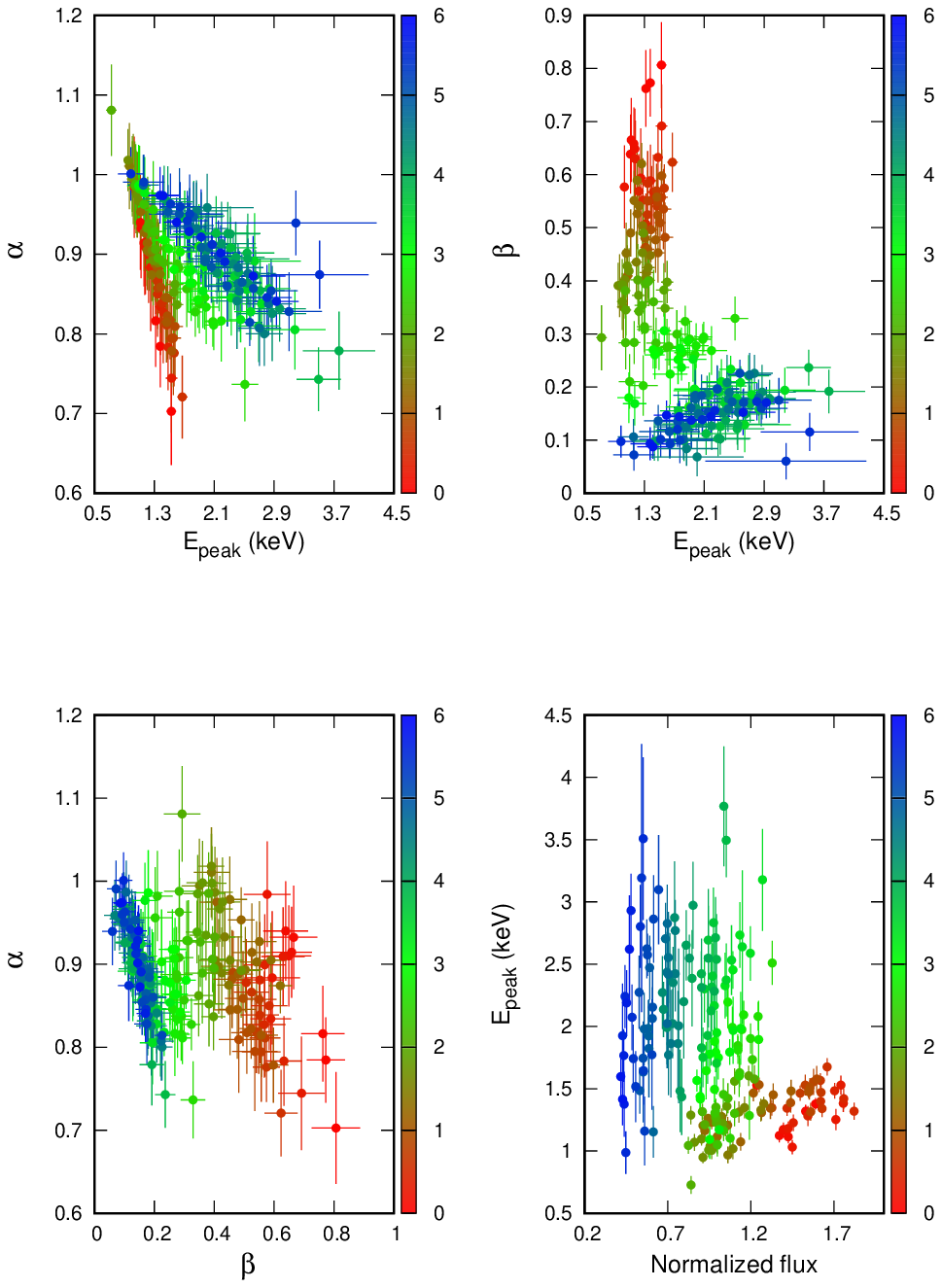} 
\caption{Correlation between the spectral parameters of the log-parabola models, which are best-fit to the simulated X-ray spectra at each snapshot of 5 ks duration for the case of gradual acceleration. The color bar indicates time evolution in days.}\label{sim_parameters_acc}
\end{figure*}

\section{Discussion}
The spectral index ($\alpha$) and the total observed flux are strongly anti-correlated at most epochs for both blazars. This result, also found by other authors \citep{kapa17,kapa18,
hota21,chandra21,wani23}, implies the ``harder-when-brighter'' nature found in blazars, i.e., spectral hardening (a decrease in absolute value of $\alpha$) with increasing flux. 
In addition, there is anti-correlation between $\alpha$ and peak energy $(E_{peak})$, which is a consequence of a higher value of peak energy when the spectrum is flatter. 

In some HSP blazars the spectral curvature ($\beta$) and peak energy ($E_{peak}$) are anti-correlated, which has been explained by energy independent acceleration probability of emitting electrons \citep{massaro06,massaro08}. The amount of energy-gain by acceleration $(\epsilon)$ is inversely proportional to the curvature ($\beta \propto  1/\text{log} (\epsilon)$) and $E_{peak} \propto \epsilon$. Hence, the inverse trend of $\beta$ and $E_{peak}$ follows. A similar trend has been observed in Mrk 421 from long-term \textit{Swift} and \textit{XMM}-Newton observations \citep{tramacere07,tramacere09,kapa17,kapa20}. However, in the individual epochs discussed here, the correlations between the variation of $\beta$ and total flux or $E_{peak}$ are weak and in some epochs the sense changes such that they are weakly correlated or anti-correlated. Therefore, over the longer timescale encompassing all epochs, there is no correlation between the above quantities.  

\citet{hota21}, have analysed the \textit{AstroSat} data from the 2017 January 3-8 epoch of Mrk 421. They fit the observed spectrum with that emitted by a log-parabolic particle distribution in the synchrotron process. They have shown that the observed flux and curvature are weakly correlated, which is consistent with our findings. \citet{rukaiya22} have found a similar trend of weak correlation for long-term Mrk 421 observations. The curvature of the spectrum may also be due to efficient stochastic acceleration of emitting electrons by magnetic turbulence at the shock front \citep{massaro11,kapa16,kapa17} in HSP blazars, if the diffusion coefficient is inversely proportional to the curvature parameter. Hence, the combined effect of statistical and stochastic acceleration process may weaken the correlation results.

\citet{massaro06,massaro08} have found a positive correlation of the spectral index $\alpha$ with curvature $\beta$. Similarly, \citet{kapa18} have obtained a strong linear relation between those parameters during a flaring event of Mrk 421 in 2016 observed by \textit{Swift}-XRT. Considering statistical acceleration, \cite{massaro04} have shown that a positive correlation occurs when the electron distribution is of log-parabola nature under the condition, in which the acceleration probability decreases with increasing particle energy. On the other hand, considering the log-parabolic particle energy distribution, \citet{hota21,rukaiya22} have found an opposite nature, i.e., a negative relation between $\alpha$ and $\beta$ during days and years-long observations of Mrk 421. During the epochs described here, $\alpha$ and $\beta$ have an anti-correlation. Hence, any clear trend in the $\alpha$-$\beta$ relationship, consistently present at epochs separated by $\sim$years, is absent, which may be caused by different physical mechanisms dominating the spectral variation at different epochs.

\begin{table}
\caption{Spearman's rank correlation coefficients between the best-fit parameters from fitting the observed spectra of 1ES 1959+650 and Mrk 421 with the log-parabola model during two epochs, with exposure times of several days, for each of the blazars.}\label{spearman_rank}
\centering
\begin{tabular}{ l c c  }
 \hline
 
 {1ES 1959+650} &3-7th Nov, 2016 & 25-26th Oct, 2017 \\
  & $r^{*}$ \quad \quad $p^{*}$ & $r$ \quad \quad $p$ \\
  \hline
 $\alpha - F_{(0.7-20 keV)}$ & -0.34 \quad \quad 0.06 & -0.97 \quad \quad 1e-7 \\
 $\alpha - E_{peak}$ & -0.94 \quad \quad 5e-15 & -0.84 \quad \quad 6e-4 \\
 $\beta - F_{(0.7-20 keV)}$ & -0.15 \quad \quad 0.4 & 0.47 \quad \quad 0.12 \\
 $\beta - E_{peak}$ & 0.44 \quad \quad 0.015 & 0.006 \quad \quad 0.98 \\
 $E_{peak} - F_{(0.7-20 keV)}$& 0.34 \quad \quad 0.06 & 0.84 \quad \quad 5e-4 \\
 $\alpha - \beta$ & -0.64 \quad \quad 1e-4 & -0.51 \quad \quad 0.09 \\
 \hline

 {Mrk 421} & 3-8th Jan, 2017 & 23-28th April, 2019 \\
  & $r$ \quad \quad $p$ & $r$ \quad \quad $p$ \\
  \hline
 $\alpha - F_{(0.7-20 keV)}$& -0.9 \quad \quad 3e-15 & -0.41 \quad \quad 0.05 \\
 $\alpha - E_{peak}$ & -0.9 \quad \quad 4e-15 & -0.41 \quad \quad 0.05 \\
 $\beta - F_{(0.7-20 keV)}$& 0.34 \quad \quad 0.03 & -0.016 \quad \quad 0.94 \\
 $\beta - E_{peak}$ & -0.27 \quad \quad 0.09 & 0.1 \quad \quad 0.62 \\
 $E_{peak} - F_{(0.7-20 keV)}$& 0.95 \quad \quad 1e-21 & 0.58 \quad \quad 0.003 \\
 $\alpha - \beta$ & -0.6 \quad \quad 3e-5 & -0.63 \quad \quad 0.001 \\
 \hline 
 
\end{tabular}
\begin{threeparttable}
\begin{tablenotes}
       \item [*] $r$ is the Spearman's correlation coefficient, $p$ denotes the probability that the null hypothesis (the quantities are uncorrelated) is true in the given sample.
\end{tablenotes}
\end{threeparttable}
\end{table}

The chromatic behavior of multi-wavelength polarization, i.e., stronger polarization at X-ray compared to longer wavelengths in HSP blazars, supports the energy stratification caused by a shock moving down the jet and energizing electrons \citep{liodakis22Natur,manel24ApJ,alan24Galax}. However, the direction of X-ray polarization w.r.t. the jet axis and its rotation at days timescale, as observed by IXPE, are sometimes not consistent with the shock-in-jet model, which indicates the presence of turbulent and/or helical magnetic field structure, as well as the contribution of other dominant processes, e.g., magnetic reconnection or stochastic acceleration \citep{marscher14,sironi14,tave2021} in the energizaton of the emitting particles. Therefore, examining detailed spectral and timing properties of jet emission, particularly from the highest energy particles, in the context of different theoretical scenarios is important.

Many authors have explored various multi-zone models of blazar emission containing different geometry and physical structure \citep{graff08,joshi11,marscher14,chen14,liu23,hu24} to understand the temporal and spectral behavior of jet emission. Different models emphasize certain aspects of jets based on the output intended for further study, e.g., SED, flux variability, polarization. For example, \citet{hota21} fit the spectrum of Mrk 421 in the 2017 Jan 3 epoch with synchrotron radiation emitted by various initial particle distributions, e.g., log-parabolic, power-law with a high-energy cut-off, system having energy-dependent diffusion and energy-dependent acceleration, etc. and studied the correlation between the best-fit parameters of the above models. They argued about the suitability of the assumed particle energy distribution models based on the results they obtained. \citet{rukaiya22} did a similar work for the long-term X-ray variation of the same blazar during 2005-2020. That is an effective approach to probe the underlying emission processes and physical parameters in the jet. Here, we study the spectral variability of two blazars in $10$ ks intervals at 3-5 epochs during 2016-19. We attempt to interpret our observational results in the context of an extensive multi-zone model, which focuses on the flux and spectral variability at hours-days timescales. In the theoretical model we use, the acceleration of the electrons may be instantaneous or gradual, the magnetic field structure of the emission region is configurable, and inverse-Compton (IC) process is also included. Although the contribution of the IC emission at the SXT and LAXPC energy bands is negligible for the sources described in this work, the energetic electrons in our model cool through both synchrotron and IC processes, which makes the resultant broadband spectrum closer to observations and hence makes our obtained inferences fairly robust. We describe the model below.

\subsection{Jet Emission Model and Its Comparison with Observed Results}
In this model, discussed in detail by \citet{aritra22,barat22}, a cylindrical emission region is assumed along the jet axis with a fixed viewing angle $(\theta)$. The emission region consists of multiple cells of equal size, having individual electron distribution and magnetic field. A shock front moves through the emission region and energizes each cell instantaneously by injection of high-energy electrons. Those electrons then cool through synchrotron and inverse-Compton processes and give rise to the emission. 

In each cell, the initial distribution of electron energy is a power law given by $N(\gamma) = N_0~\gamma^{-s}$, where $\gamma$ is the Lorentz factor of electrons with lower and upper limits $\gamma_{min}$ and $\gamma_{max}$, respectively, while $s$ and $N_0$ are constants. The electron energy evolves as $d\gamma/dt= -k_2 \gamma^2$ due to radiative cooling. Here $k_2 = \frac{4 \sigma_T}{3 m_e c} (U_B+U_R)$, $\sigma_T$ is the Thomson scattering cross-section, $U_B$ is the magnetic energy density and $U_R$ is the energy density of seed photons for SSC radiation. To simulate X-ray emission by the synchrotron process as is the case for the two HSP blazars discussed here, we use a set of input parameters for our model, which are listed in Table \ref{theo_para}. We simulate light curves at soft ($0.7-7$ keV) and hard ($7-20$ keV) X-ray energies of duration $\sim$5 days which are corrected for light-travel time effect to reach the observer's frame. We bin the simulated data in intervals of $5$ ks. We use shorter time bins compared to those used in the observed data to facilitate the identification of the trends in the variation of the best-fit spectral parameters such as those shown in, e.g., Figure \ref{sim_parameters_acc} and to avoid the influence of spurious changes, if any, in our interpretation. It will also help in comparing those results with other observed data sets, in which use of shorter time bins is possible. Then we fit the $0.7-20$ keV spectrum at each interval with a log-parabola function. Thus, we obtain the temporal variation of the best-fit parameters similar to what we calculated from the observed data. In \citet{das23}, we found both zero and non-zero (hard/soft) time lags between the hard and soft X-ray light curves during the above epochs. The non-zero lags can be explained by gradual acceleration and synchrotron cooling processes. Therefore, in order to generate X-ray light curves with similar temporal properties, we consider two cases, e.g., one in which the synchrotron cooling timescale of the highest energy electrons dominates over the acceleration timescale giving rise to soft lags and another in which the acceleration timescale dominates causing hard lag. We discuss those two cases below.

\subsubsection{Case-I: Radiative Cooling}
When radiative cooling dominates, we fix $\gamma_{max}$ at a high value ($10^5-10^6$) in each cell such that the synchrotron peak is at keV energies, as observed in the blazars in our sample. The simulated light curves and spectra are shown in Figure \ref{sim_lc_sed_noacc}. Figure \ref{sim_parameters_noacc} shows the correlations between pairs of spectral parameters of the best-fit log-parabola models at individual time-bins. It shows that $\alpha$ has a strong anti-correlation with the peak energy $E_{peak}$, which is similar to what we found in the observed data in most epochs. At the beginning, i.e., soon after the energy injection, the peak energy ($\gamma_{max}$) of individual cells is high and consequently the power-law index of the emitted spectrum is flatter while the emitted flux is also high. As time passes, the electron energy distribution in the cells becomes steeper (magnitude of $s$ becomes larger) causing the emitted spectrum to become steeper (value of $\alpha$ becomes higher). This happens because higher-energy electrons cool faster through synchrotron radiation, e.g., $t_{cool}\sim \gamma^{-1}$, which causes the relative fraction of higher-energy electrons to decrease with time. This is accompanied by an overall decrease in the emitted flux because the emitting electrons lose energy through radiative cooling. That causes the $\alpha$-$E_{peak}$ anti-correlation in the theoretical result such as that exhibited in Figure \ref{sim_parameters_noacc} top left panel. In the case of $\beta$ vs $E_{peak}$, the simulated data shows correlation in some sub-intervals, denoted by different colors in Figure \ref{sim_parameters_noacc} top right panel. However, the overall trend considering the entire time span ($\sim$5 days) of the simulation is of weak anti-correlation. Consequently, the variation of $\alpha$ and $\beta$ exhibits weak anti-correlation in some sub-intervals while there is a weak correlation over the total time span, as shown in Figure \ref{sim_parameters_noacc} bottom left panel.
On the other hand, the flux and peak energy are correlated in the simulation as shown in Figure \ref{sim_parameters_noacc} bottom right panel, which is consistent with the observed harder-when-brighter trend.

\subsubsection{Case-II: Gradual Acceleration and Radiative Cooling}
In this case, we implement gradual acceleration by increasing the value of $\gamma_{max}$ in successive cells along the emission region. 
The simulated light curves and spectra are shown in Figure \ref{sim_lc_sed_acc}. After fitting the spectra, we obtain the correlations of parameters shown in Figure \ref{sim_parameters_acc}. $\alpha$ has a weaker anti-correlation with $E_{peak}$ compared to the cooling-dominated case. Multiple values of $\alpha$ are obtained for the same value of $E_{peak}$ and vice versa as shown in Figure \ref{sim_parameters_acc} top left panel. That is not commonly seen in the observed results but is indeed found in some epochs, e.g., in Mrk 421 during the 2019 Apr 23 epoch. $\beta$ has weak correlation or anti-correlation at different intervals, with $E_{peak}$, similar to the observed results. 
$\alpha$ and $\beta$ show anti-correlation at successive intervals denoted by different colors in Figure \ref{sim_parameters_acc} bottom left panel. The overall $\alpha$-$\beta$ anti-correlation considering the total 5-day time span is weak. This behavior is similar to the observed results, e.g., bottom right panels of Figures \ref{para_log_1959} and \ref{para_log_421}. The flux and peak energy, on the other hand, are uncorrelated, which is different from the observed correlation among those parameters in Mrk 421. In 1ES 1959+650, however, the low-flux states do not exhibit the said correlation between flux and E$_{peak}$, which matches with the above simulated result. When gradual acceleration is introduced in the theoretical model, the synchrotron peak energy increases with time but the total flux may decrease because electrons lose energy through radiation, the combined effect of which may be responsible for the lack of correlation noted above.

\begin{table}
\begin{center}
\caption{Fixed parameter values used in the theoretical model of jet emission.}\label{theo_para}
\begin{tabular}{ l c  }
 \hline
 Model Parameter & Value \\
 \hline
 Bulk Lorentz factor ($\Gamma$) & 15  \\  
 Doppler factor ($\delta$) & 20 \\
 Viewing angle with jet axis ($\theta$) in degree & 2.7 \\
 Slope of electron distribution ($s$) & $2.1$ \\
 Minimum Lorentz factor of electron ($\gamma_{min}$) & $100$\\
 Magnetic field in the first cell ($B_i$) in G & $0.2$\\
 Magnetic field in the last cell ($B_f$) in G & $0.1$\\ 
  \hline
\end{tabular}
\end{center}
\end{table}
\medskip
\medskip

\section{Summary \& Conclusion}
In this paper, we have analysed the X-ray spectra of two HSP blazars Mrk 421 and 1ES 1959+650 at multiple epochs during 2016-19 at $0.7-20$ keV using \textit{AstroSat} and \textit{Swift} data. We have fitted the average spectrum at each epoch with the log-parabola function. We carried out time-resolved spectroscopy with bins of length 10 ks to find the temporal variation of the best-fit spectral parameters of the log-parabola model. Finally, we compared the results of our analyses to that from two cases of a theoretical model of blazar jet emission, namely, one in which the acceleration timescale dominates over the radiative cooling timescale of the highest energy electrons and \textit{vice versa}. In that model, high-energy particles with a simple power-law energy distribution are injected, at a timescale shorter or longer than the cooling timescale of the highest-energy electrons, in the jet emission region by a moving shock front. The particles cool via radiation through synchrotron and inverse-Compton processes in the presence of a turbulent magnetic field in the jet. The curvature of the simulated X-ray spectrum is driven by the radiative cooling while the flux and spectral variability are primarily due to the injection of high-energy particles by the moving front and the fluctuating magnetic field. We find that:\\
1. The time-resolved spectroscopy revealed a strong anti-correlation of the spectral index $\alpha$ with the total flux and peak energy $E_{peak}$, which indicates a harder-when-brighter trend in the blazars. Furthermore, the correlations of the curvature parameter $\beta$ with flux and $E_{peak}$ are weak and change sense from one epoch to another, indicating no overall trend while there is an anti-correlation between $\alpha$ and $\beta$.\\
2. The strong anti-correlation of $\alpha$ and $E_{peak}$, deduced from the observed spectral variability of Mrk 421 and 1ES 1959+650, match well with the cooling-dominated case of our theoretical model. Similarly, the general harder-when-brighter trend of the X-ray variation is also much better reproduced in the cooling dominated scenario. On the other hand, $\alpha$ and $\beta$ exhibit a strong anti-correlation in Mrk 421, which is better reproduced in the acceleration dominated case of the model. In one epoch each in Mrk 421 (2019 Apr 23) and 1ES 1959+650 (2017 Oct 25) the nature of the $\alpha$-$E_{peak}$ correlation is different, to a certain extent, from the rest and consequently multiple values of $E_{peak}$ can be found for the same value of $\alpha$ and vice versa (bottom left panels of Figures \ref{para_log_1959} and \ref{para_log_421}). That is also found in the simulated results of the acceleration dominated case. Therefore, both cases with gradual acceleration and radiative cooling are partially able to reproduce observational results. This implies that the basic features of the model, i.e., high-energy particles, having a simple power-law energy distribution injected by a shock front moving down the emission region, emitting synchrotron and IC radiation in the presence of a fluctuating magnetic field can give rise to the observed spectral and flux variability properties at hours to days timescales. However, a model with more complex features in the energy distribution of injected particles, geometric aspects, and emission processes as well as more detailed acceleration mechanisms, e.g., including stochastic acceleration are needed for a better match with all of the observed parameter trends and consequently a more comprehensive interpretation of the observed results. \\
3. Application of a flexible and extensive theoretical model to interpret the comprehensive data set used here makes our inferences fairly robust. It also has the scope of providing more general insight into the blazar phenomena by increasing the number of objects in the sample by a factor of a few and/or extending the parameter space of the theoretical model in a future work.

\section{Data Availability}
\textit{AstroSat} data are available in its archival data repository.\\
Weblink:\url{https://astrobrowse.issdc.gov.in/astro_archive/archive/Home.jsp}\\
The software for data reduction of \textit{AstroSat} are available in \textit{AstroSat Science Support Cell (ASSC)}. \\
Weblink:\url{http://astrosat-ssc.iucaa.in}\\
The \textit{Swift} data are available in NASA's \textit{HEASARC} archive repository.\\
Weblink:\url{https://heasarc.gsfc.nasa.gov/}\\
The \textit{Swift}-XRT products are generated by online XRT building product tool.\\
Weblink:\url{https://www.swift.ac.uk/user_objects/index.php}

\section{Acknowledgment}
We thank the anonymous referee for their comments, which have made this work more comprehensive and have provided additional clarity to the presentation. This work has used data from the Indian Space Science Data Centre (ISSDC) under the \textit{AstroSat} mission of the Indian Space Research Organisation (ISRO). We acknowledge the POC teams of the SXT and LAXPC instruments for archiving data and providing the necessary software tools. The data of \textit{Swift}-XRT has been taken from NASA's data repository High Energy Astrophysics Science Archive Research Center (HEASARC). We acknowledge \textit{Swift} team for providing the online software analysis tool. We thank ISRO for support under the \textit{AstroSat} archival data utilization program, ANRF for a SURE grant (SUR/2022/001503), and IUCAA for their hospitality and usage of their facilities during our stay at different times as part of the university associateship program. RC thanks Presidency University for support under the Faculty Research and Professional Development (FRPDF) Grant. We are thankful to Kaustav Mitra and Arit Bala for developing the numerical jet code, which has been used here. We thank Sunil Chandra, Ritesh Ghosh, Savithri Ezhikode, Gulab Dewangan and Ranjeev Misra for fruitful discussion regarding data analysis for this work. \\ \\ \\ \\

\bibliographystyle{mnras}
\bibliography{spectral_variability}

\appendix

\section{Spectral Modeling}
The log-parabola model in terms of photon flux can be expressed as: 
 \begin{equation}
 F(E) = N_0 (E/E_0)^{-\alpha -\beta log_{10}(E/E_0)} ~~~~~~~
 \end{equation}
 where $N_0$ is a normalization factor, $E_0$ is pivot energy, and $\alpha$ and $\beta$ are the photon index and curvature around the peak of the function, respectively. Using those parameters, the peak energy and height of the peak are calculated by, $E_p = E_0 10^{(2-\alpha)/2\beta}$ keV and $S_p = E_p^2 F(E_p)$ ~erg\,cm$^{-2}$\,s$^{-1}$ \citep{massaro08}.
 
\begin{figure*}
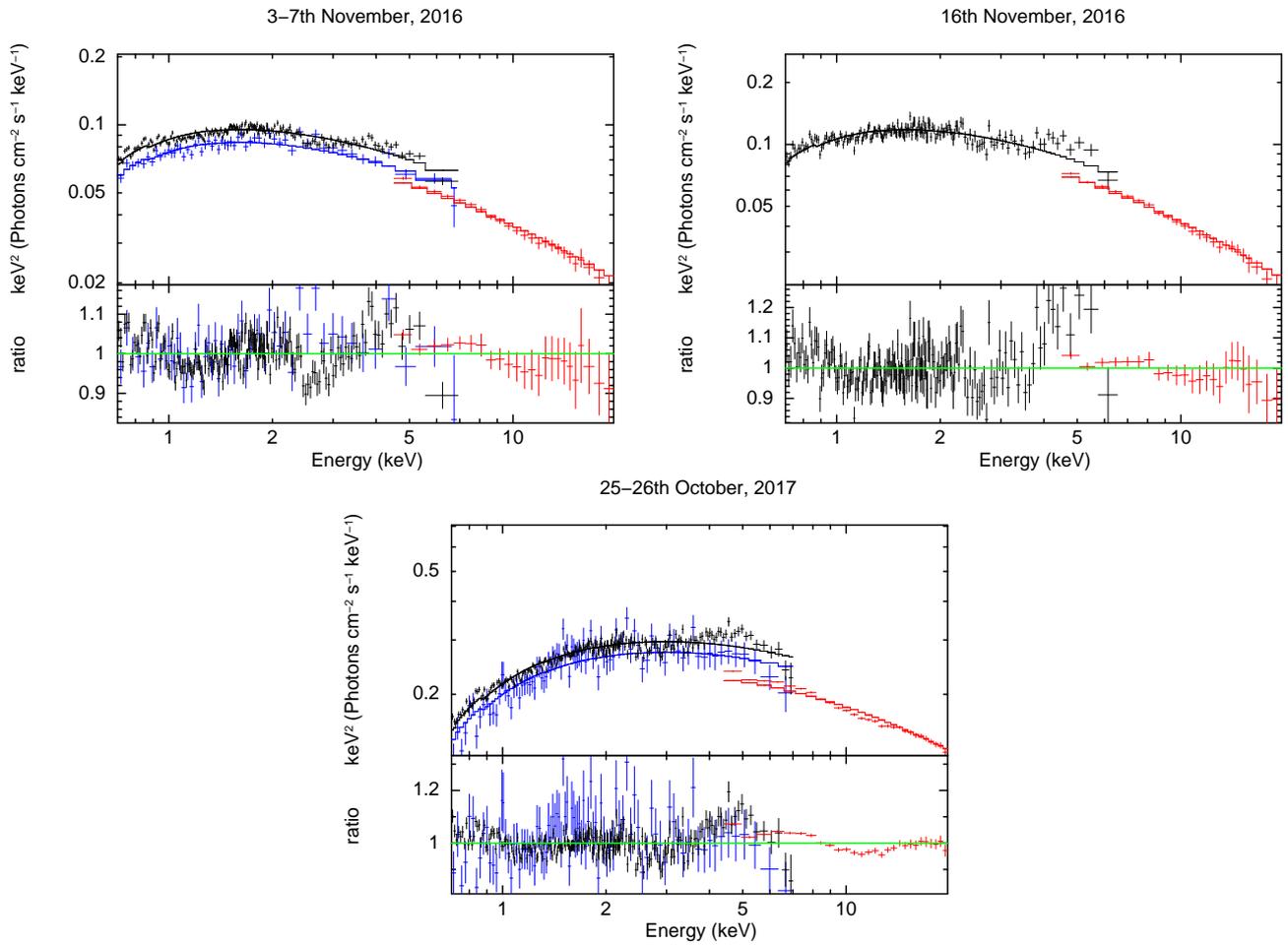

   \begin{minipage}{0.5\textwidth}
     \centering
     \includegraphics[width=.7\linewidth,angle=-90]{figures/1959_xrt_sxt_lxp_0.7-20_2016.eps}
     \end{minipage}\hfill
     \begin{minipage}{0.5\textwidth}
     \centering
     \includegraphics[width=.7\linewidth,angle=-90]{figures/sxt_lxp_spec1959_0800_0.7-20.eps} 
   \end{minipage}
   \begin{minipage}{0.5\textwidth}
     \centering
     \includegraphics[width=.7\linewidth,angle=-90]{figures/1959_xrt_sxt_lxp_0.7-20_2017.eps}
   \end{minipage}
   \caption{XRT (blue), SXT (black) and LAXPC (red) data points and the best-fit log-parabola model (solid lines) of the combined unfolded spectra of the blazar 1ES 1959+650 during three epochs in 2016-17.}\label{sed_logpar_1959}
\end{figure*}

\begin{figure*}
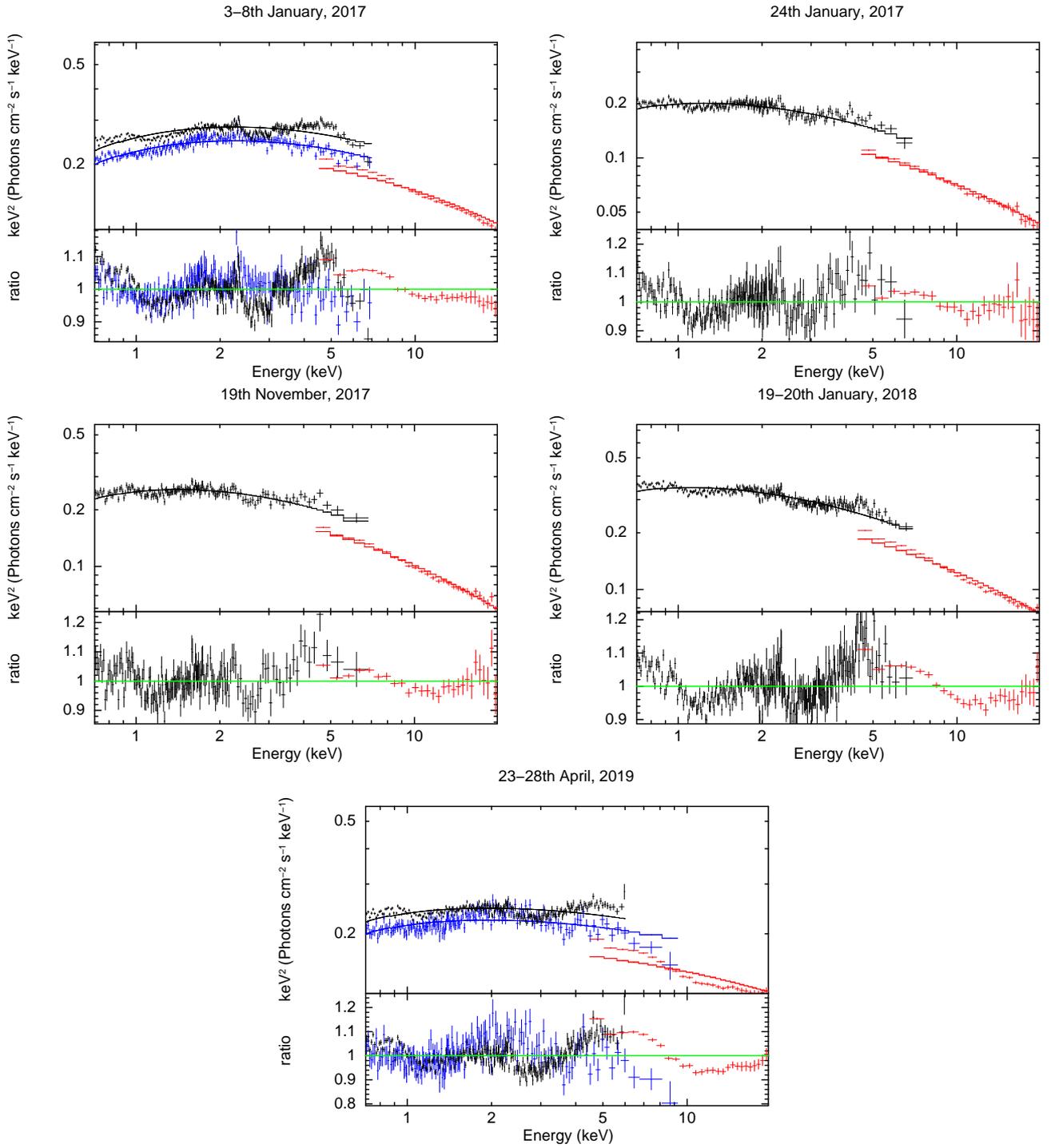

   \begin{minipage}{0.5\textwidth}
     \centering
     \includegraphics[width=.7\linewidth,angle=-90]{figures/421_xrt_sxt_lxp_0.7-20_2017.eps}
   \end{minipage}\hfill
      \begin{minipage}{0.5\textwidth}
     \centering
     \includegraphics[width=.7\linewidth,angle=-90]{figures/sxt_lxp_spec421_0978_0.7-20.eps}
   \end{minipage}\hfill
      \begin{minipage}{0.5\textwidth}
     \centering
     \includegraphics[width=.7\linewidth,angle=-90]{figures/sxt_lxp_spec421_1704_0.7-20.eps}
   \end{minipage}\hfill
      \begin{minipage}{0.5\textwidth}
     \centering
     \includegraphics[width=.7\linewidth,angle=-90]{figures/sxt_lxp_spec421_1852_0.7-20.eps}
   \end{minipage}\hfill
   \begin{minipage}{0.5\textwidth}
     \centering
     \includegraphics[width=.7\linewidth,angle=-90]{figures/421_xrt_sxt_lxp_0.7-20_2019.eps} 
   \end{minipage}
   \caption{XRT (blue), SXT (black) and LAXPC (red) data points and the best-fit log-parabola model (solid lines) of the combined unfolded- spectra of the blazar Mrk 421 during five epochs in 2017-19.}\label{sed_logpar_421}
\end{figure*}

The average spectra are fitted by the above model for all epochs of Mrk 421 and 1ES 1959+650. The best-fit models are shown in Figures \ref{sed_logpar_1959} and \ref{sed_logpar_421}.
 
\label{lastpage}
\end{document}